\documentclass[10pt]{article}
\usepackage{latexsym}
\usepackage{amssymb}
\usepackage{amsmath}
\usepackage{amscd}
\usepackage{amsthm}
\usepackage[left=2cm,top=2.5cm,right=2.5cm,bottom=1.5cm]{geometry}

\usepackage[dvips]{graphicx}
\usepackage{hyperref}
\begin{document}
\begin{center}
\Large{\bf{Lyra's Cosmology of Massive String in Anisotropic Bianchi-II Space-time }}
\\
%\vspace{10mm} \normalsize{} \vspace{5mm}
\vspace{10mm} \normalsize{Anil Kumar Yadav$^\dag$ and Abdul Haque$^\ddag$}\\ \vspace{4mm} 
\normalsize{$^\dag$Department of Physics, Anand Engineering
College, Keetham, Agra-282 007, India} \\
\vspace{2mm}
\normalsize{E-mail: abanilyadav@yahoo.co.in}\\
\vspace{4mm}
\normalsize{$^\ddag$Department of Mathematics, P. G. College, Ghazipur - 233 001, India}\\
\vspace{2mm}
\normalsize{E-mail: abdulg\_haque@rediffmail.com}

\end{center}

\begin{abstract}
The paper deals with a spatially homogeneous and totally anisotropic Bianchi II cosmological models 
representing massive strings in normal gauge for Lyra's manifold. The modified Einstein's field equations 
have been solved by applying variation law for Hubble's parameter. This law generates two type of solutions 
for average scale factor, one is of power law type and other is of exponential law type. 
The power law describes the dynamics of Universe from big bang to present epoch while exponential 
law seems reasonable to project dynamics of future Universe. It has been found that the displacement actor $(\beta)$ 
is a decreasing function of time and it approaches to small positive value at late time, which is 
collaborated with Halford (1970) as well as recent observations of SN Ia. The study reveals that 
massive strings dominate in early Universe and eventually disappear from Universe for sufficiently large time,
 which is in agreement with the current
astronomical observations.
\end{abstract}
\smallskip


%\newpage
%%%%%%%%%%%%%%%%%%%%%%%%%%%%%%%%%%%%%%%%%%%%%%%%%%%%%%%%%%%%%%%%%%%%%%%%%%%%%%%%%%%%%%%%%%%%%%%%%%%%%
%%%%%%%%%%%%%%%%%%%%%%%%%%%%%%%   SECTION 1  %%%%%%%%%%%%%%%%%%%%%%%%%%%%%%%%%%%%%%%%%%%%%%%%%%%%%%%%%%
\section{Introduction}
In recent years, there has been considerable interest in string
cosmology because cosmic strings play an important role in the study of early Universe. 
It is generally assumed that after the big bang, 
the Universe may have undergone a series of phase transitions as its temperature lowered down below some 
critical temperature as predicted by grand unified theories \cite{ref1}$-$\cite{ref6}. 
At the very early stages of evolution  
of the Universe, it is believed that during phase transition the symmetry of the universe is broken spontaneously. 
It can give rise to topologically stable defects such as domain walls, strings and monopoles. 
In particular, cosmic strings are produced in the breaking of U(1) symmetry, are good candidates to seed the 
formation of galaxies. These cosmic strings have
stress-energy, and couple to the gravitational field. Therefore, it
is interesting to study the gravitational effects that arise from
strings. The pioneering work in the formulation of the
energy-momentum tensor for classical massive strings was done by
Letelier \cite{ref7} who considered the massive strings to be formed
by geometric strings with particle attached along its extension.
Letelier \cite{ref8} first used this idea in obtaining cosmological
solutions in Bianchi-I and Kantowski-Sachs space-times. Stachel
\cite{ref9} has studied massive string.

The observed Universe is satisfactorily described by homogeneous
and isotropic models given by the FRW space-time. But at smaller
scales, the Universe is neither homogeneous and isotropic nor do we
expect the Universe in its early stages to have these properties. The anomalies found 
in the cosmic microwave background (CMB) and large scale structure 
observations stimulated a growing interest in anisotropic cosmological model of Universe. 
Here we confine ourselves to models of Bianchi-type II. 
Bianchi type-II space-time
has a fundamental role in constructing cosmological models suitable
for describing the early stages of evolution of Universe. Asseo and Sol \cite{ref10} emphasized the
importance of Bianchi type-II Universe. 

Roy and Banerjee \cite{ref11} have dealt with locally rotationally
symmetric (LRS) cosmological models of Bianchi type-II representing
clouds of geometrical as well as massive strings. Wang \cite{ref12}
studied the Letelier model in the context of LRS Bianchi type-II
space-time. Recently, Pradhan et al. \cite{ref13,ref14} and
Amirhashchi and Zainuddin \cite{ref15} obtained LRS Bianchi type II
cosmological models with perfect fluid distribution of matter and
string dust respectively. Belinchon \cite{ref16,ref17} studied
Bianchi type-II space-time in connection with massive cosmic string
and perfect fluid models with time varying constants under the
self-similarity approach respectively. Recently, Kumar
\cite{ref18} has investigated string cosmological models in Bianchi
type-II space-time.

In last few decades there has been considerable interest in alternative theories of gravitation. 
The most important among them being scalar-tensor theories proposed by Lyra
\cite{ref19}. Lyra suggested a modification of Riemannian geometry which
may also be considered as a modification of Wey's geometry. In Lyra's geometry, Weyl's
concept of gauge, which is essentially a metrical concept, is modified by the introduction
of a gauge function into the structure less manifold.
This alternating theory is of interest because it produces effects similar to those produced
in Einstein's theory. Also vector field in this theory plays similar role to cosmological
constant in general relativity \cite{ref20,ref21}. Several authors Sen
and Vanstone \cite{ref22}, Bhamra \cite{ref23}, Singh and Singh \cite{ref24}, Rahaman et al \cite{ref25, ref26}, 
Pradhan et al \cite{ref27}$-$\cite{ref31}, Yadav \cite{ref32} and 
Yadav et al \cite{ref33,ref34} 
have studied cosmological models
based on Lyra's manifold in various physical contexts.\\

Recently, Agwarwal, Pandey and Pradhan \cite{ref35} have studied Bianchi II string cosmological model 
in Lyra's manifold. They have investigated time varying vector field by assuming direction of string along 
x-axis. In this paper, we have assumed the direction of string along z-axis, collaborated with Saha \cite{ref36}.
 The paper is organized as follows. The
metric and the field equations are presented in Section 2. Section 3
deals with exact solutions of the field equations with
strings fluid. Physical behavior of the derived models are elaborated in
detail. Finally, in Section 4, concluding remarks are given.

%%%%%%%%%%%%%%%%%%%%%%%%%%%%%%%%%%%%%%%%%%%%%%%%%%%%%%%%%%%%%%%%%%%%%%%%%%%%%%%%%%%%%%
%%%%%%%%%%%%%%%%%%%%%%%%%%%%%%%  SECTION 2  %%%%%%%%%%%%%%%%%%%%%%%%%%%%%%%%%%%%%%%%%
\section{The metric and field  equations}
We consider totally anisotropic Bianchi type-II line element, given by
\begin{equation}
\label{eq1} ds^{2} = - dt^{2} + A^{2}(dx - zdy)^{2} + B^{2} dy^{2} +
C^{2} dz^{2},
\end{equation}
where the metric potentials $A$, $B$ and $C$ are functions of $t$
alone. This ensures that the model is spatially homogeneous.

The field equations ( in gravitational units $c = 1, 8\pi
G = 1 $), in normal gauge for Lyra's manifold, obtained by Sen \cite{ref37} as
\begin{equation}\label{eq2}
 R_{i}^{j}-\frac{1}{2}g_{i}^{j}R +\frac{3}{2}\phi_{i}\phi^{j}-\frac{3}{4}g_{i}^{j}\phi_{k}\phi^{k}=  -
T^{i}_{j},
\end{equation}
where $\phi_{i}$ is the is the displacement field vector defined as
\begin{equation}
\label{eq3}
\phi_{i}=(0,0,0,\beta(t))
\end{equation}
and the other symbols have their usual meaning as in Riemannian geometry.\\  
Einstein tensor. The energy-momentum tensor $T^{i}_{j}$ for a cloud
of massive strings and perfect fluid distribution is taken as
\begin{equation}
\label{eq4} T^{i}_{j} = (\rho + p)v^{i}v_{j} + p g^{i}_{j} -\lambda
x^{i}x_{j},
\end{equation}
where $p$ is the isotropic pressure; $\rho$ is the proper energy density for a cloud strings with particles
attached to them; $\lambda$ is the string tension density; $v^{i}=(0,0,0,1)$ is the four-velocity of the
particles, and $x^{i}$ is a unit space-like vector representing the direction of string. The vectors $v^{i}$
and $x^{i}$ satisfy the conditions
\begin{equation}
\label{eq5} v_{i}v^{i}=-x_{i}x^{i}=-1,\;\; v^{i}x_{i}=0.
\end{equation}

Choosing $x^{i}$ parallel to $\partial/\partial z$, we have
\begin{equation}
\label{eq6} x^{i} = (0,0,C^{-1},0).
\end{equation}

Here the cosmic string has been directed along z-direction in order
to satisfy the condition $T^{1}_{1}=T^{2}_{2}$. As a result, the
off-diagonal component of Einstein tensor, viz.,
$G^{1}_{2}=z(T^{2}_{2}-T^{1}_{1})$ vanishes. A detailed analysis
about the choice of energy-momentum tensor for Bianchi type-II
models is given by Saha \cite{ref36}.

If the particle density of the configuration is denoted by
$\rho_{p}$, then
\begin{equation}
\label{eq7} \rho = \rho_{p}+\lambda.
\end{equation}

The modified Einstein's field equations (\ref{eq2}) for line element
(\ref{eq1}) with energy-momentum tensor (\ref{eq4}), lead to the
following set of independent differential equations:
\begin{equation}
\label{eq8} \frac{\ddot{B}}{B} + \frac{\ddot{C}}{C} +
\frac{\dot{B}\dot{C}}{BC} - \frac{3}{4}\frac{A^{2}} {B^{2}C^{2}} - \frac{3}{4}\beta^{2} =
-p \;,
\end{equation}
\begin{equation}
\label{eq9} \frac{\ddot{C}}{C} + \frac{\ddot{A}}{A} + \frac{\dot{C}\dot{A}}{CA} + \frac{1}{4}\frac{A^{2}}
{B^{2}C^{2}} - \frac{3}{4}\beta^{2} = -p\;,
\end{equation}
\begin{equation}
\label{eq10} \frac{\ddot{A}}{A} + \frac{\ddot{B}}{B} +
\frac{\dot{A}\dot{B}}{AB} + \frac{1}{4}\frac{A^{2}} {B^{2}C^{2}}- \frac{3}{4}\beta^{2} =
-p+ \lambda\;,
\end{equation}
\begin{equation}
\label{eq11} \frac{\dot{A}\dot{B}}{AB} + \frac{\dot{B}\dot{C}}{BC} + \frac{\dot{C}\dot{A}}{CA} -
\frac{1}{4}\frac{A^{2}}{B^{2}C^{2}}+ \frac{3}{4}\beta^{2} = \rho\;.
\end{equation}

Here, and in what follows, an over dot indicates ordinary
differentiation with respect to $t$. The energy conservation
equation $\;T^{ij}_{\;\;\;;j}=0$, leads to the following expression:
\begin{equation}
\label{eq12} \dot\rho + (\rho + p)\left(\frac{\dot{A}}{A} +
\frac{\dot{B}}{B} + \frac{\dot{C}}{C}\right) -
\lambda\frac{\dot{C}}{C} = 0\;,
\end{equation}
and conservation of R. H. S. of eq. (\ref{eq2}) leads to
\begin{equation}
 \label{eq13}
\left(R_{i}^{j}-\frac{1}{2}g_{i}^{j}R\right)_{ ;j} +\frac{3}{2}\left(\phi_{i}\phi^{j}\right)_{ ;j} - \frac{3}{4}
\left(g_{i}^{j}\phi_{k}\phi^{k}\right)_{ ;j} = 0
\end{equation}
Equation (\ref{eq13}) reduces to
\begin{eqnarray}\label{eq14}
\frac{3}{2}\phi_{i}\left[\frac{\partial\phi^{j}}{\partial x^{j}}+\phi^{l}\Gamma^{j}_{lj}\right] +\frac{3}{2}
\left[\frac{\partial\phi_{i}}{\partial x^{j}}-\phi_{l}\Gamma^{l}_{ij}\right] -\frac{3}{4}g_{i}^{j}\phi_{k}
\left[\frac{\partial\phi^{k}}{\partial x^{j}} + \phi^{l}\Gamma_{lj}^{k}\right]
-\frac{3}{4}g_{i}^{j}\phi^{k}\left[\frac{\partial\phi_{k}}{\partial x^{j}} - \phi_{l}\Gamma_{kj}^{l}\right] = 0
\end{eqnarray}
Equation (\ref{eq14}) is identically satisfied for $i = 1, 2, 3$. for $i = 4$, eq. (\ref{eq14}) reduces to
\begin{eqnarray}
\label{eq15}
\frac{3}{2}\beta\left[\frac{\partial(g^{44}\phi_{4})}{\partial x^{4}}+\phi^{4}\Gamma^{4}_{44}\right] 
+\frac{3}{2}g^{44}\phi_{4}\left[\frac{\partial\phi_{4}}{\partial t}-\phi_{4}\Gamma^{4}_{44}\right] - 
\frac{3}{4}g^{4}_{4}\phi_{4}\left[\frac{\partial\phi^{4}}{\partial x^{4}}+\phi^{4}\Gamma^{4}_{44}\right] \nonumber\\
 - \frac{3}{4}g^{4}_{4}g^{44}\phi^{4}\left[\frac{\partial\phi_{4}}{\partial t}-\phi^{4}\Gamma^{4}_{44}\right] = 0
\end{eqnarray}
which leads to
\begin{equation}
\label{eq16}
\frac{3}{2}\beta\dot{\beta} + \frac{3}{2}\beta^{2}\left(\frac{\dot{A}}{A}+\frac{\dot{B}}{B}
+\frac{\dot{C}}{C}\right)=0
\end{equation}
Thus eq. (\ref{eq12}) combined with eq. (\ref{eq16}) is the resulting equation when energy conservation equation 
is satisfied in the given system.\\

%%%%%%%%%%%%%%%%%%%%%%%%%%%%%%%%%%%%%%%%%%%%%%%%%%%%%%%%%%%%%%%%%%%%%%%%%%%%%%%%%%%%%%%%%%%%%%%%%%
%%%%%%%%%%%%%%%%%%%%%%%%%%%%%%%  SECTION 3  %%%%%%%%%%%%%%%%%%%%%%%%%%%%%%%%%%%%%%%%%%%%%%%%%%%%%
\section{Solutions of the Field Equations}
Equations (\ref{eq8})-(\ref{eq11}) and (\ref{eq16})are five equations in seven unknown
parameters $A$, $B$, $C$, $p$, $\rho$, $\lambda$ and $\beta$. Two additional
constraints relating these parameters are required to obtain
explicit solutions of the system.

First, we utilize the special law of variation for the Hubble's
parameter given by Berman \cite{ref38}, which yields a constant
value of deceleration parameter. Here, the law reads as
\begin{equation}
\label{eq17} H = \ell (ABC)^{-\frac{n}{3}},
\end{equation}
where  $\ell>0$  and  $n\geq0$ are constants. Such type of relations
have already been considered by Berman and Gomide \cite{ref39} for
solving FRW models. Later on, many authors (see Kumar and Singh \cite{ref40}, Pradhan et al \cite{ref41}, 
Yadav \cite{ref42} 
and references therein)
have studied Bianchi type models by using the special
law for Hubble's parameter that yields constant value of
deceleration parameter.

Considering $(ABC)^{\frac{1}{3}}$ as the average scale factor of the
anisotropic Bianchi-II space-time, the average Hubble's parameter
may be written as
\begin{equation}
\label{eq18} H = \frac{1}{3}\left(\frac{\dot{A}}{A} +
\frac{\dot{B}}{B} + \frac{\dot{C}}{C}\right).
\end{equation}

Equating the right hand sides of (\ref{eq13}) and (\ref{eq14}), and
integrating, we obtain
\begin{equation}
\label{eq19} ABC = (n \ell t + c_{1})^{\frac{3}{n}}, \;\;\;\;(n\neq
0)
\end{equation}
\begin{equation}
\label{eq20} ABC = c_{2}^{3}e^{3 \ell t},\;\;\;\;(n=0)
\end{equation}
where $c_{1}$ and $c_{2}$ are constants of integration. Thus, the
law (\ref{eq17}) provides power-law (\ref{eq19}) and exponential-law
(\ref{eq20}) of expansion of the Universe.

Following Pradhan and Chouhan \cite{ref41}, we assume that the
component $\sigma^{1}_{~1}$ of the shear tensor $\sigma^{j}_{~i}$ is
proportional to the expansion scalar ($\theta$). This condition
leads to the following relation between the metric potentials:
\begin{equation}
\label{eq21} A = (BC)^{m},
\end{equation}
where $m$ is a positive constant.

Now, subtracting (\ref{eq9}) from (\ref{eq8}), and taking integral
of the resulting equation two times, we get
\begin{equation}
\label{eq22} \frac{B}{A} = c_{4} \exp \left[\int
\left\{\frac{1}{ABC}\int\frac{ A^{3}}{BC}dt\right\} dt+c_{3} \int
\frac{1}{ABC} dt \right],
\end{equation}
where  $\;c_{3} $ and  $\;c_{4} $ are constants of integration. In
the following subsections, we discuss the string cosmology using the
power-law (\ref{eq19}) and exponential-law (\ref{eq20}) of expansion
of the Universe.
%%%%%%%%%%%%%%%%%%%%%%%%%%%%%%%%%%%%%%%%%%%%%%%%%%%%%%%%%%%%%%%%%%%%%%%%%%%%%%%%%%%%%%%%%%%%%%%%%%%%%%
%%%%%%%%%%%%%%%%%%%%%%%%%%%%%%%%%%%%%%%%%%%%%%%%%%%%%% SUBSECTION 3.1 %%%%%%%%%%%%%%%%%%%%%%%%%%%%%%%%%
\subsection{Lyra's Cosmology of String fluid with Power-law}
Solving the equations (\ref{eq16}), (\ref{eq12}) and (\ref{eq19}),
we obtain the metric functions as
\begin{equation}
\label{eq23} A(t) = (n \ell t + c_{1})^{\frac{3m}{n(m + 1)}} \;,
\end{equation}
\begin{equation}
\label{eq24} B(t) = c_{4}(n \ell t + c_{1})^{\frac{3m}{n(m + 1)}}
\exp{\left[\frac{(m+1)^{2}}{2\ell ^{2}M}(n\ell t +
c_{1})^{\frac{6(m-1)}{n(m+1)}+2 }+\frac{c_{3} }{ \ell (n-3)} (n\ell
t + c_{1})^{\frac{n - 3}{n} }\right]} \;,
\end{equation}
\begin{equation}
\label{eq25} C(t) = c_{4}^{-1}(n \ell t + c_{1})^{\frac{3(1-m)}{n(m
+ 1)}} \exp{\left[-\frac{(m+1)^{2}}{2\ell ^{2}M}(n\ell t +
c_{1})^{\frac{6(m-1)}{n(m+1)}+2 }-\frac{c_{3} }{ \ell (n-3)} (n\ell
t + c_{1})^{\frac{n - 3}{n} }\right]} \;,
\end{equation}
where $M=(9m-3+mn+n)(3m-3+mn+n)$ and $\;n\neq3$.
Thus, the metric (\ref{eq1}) is completely determined.\\
Equation (\ref{eq16}) leads to
\begin{equation}
\label{eq26}
\frac{\dot{\beta}}{\beta}=-\left(\frac{\dot{A}}{A}+\frac{\dot{B}}{B}+\frac{\dot{C}}{C}\right)
\end{equation}
which reduces to
\begin{equation}
\label{eq27}
\frac{\dot{\beta}}{\beta}=-\frac{3l}{(n\ell t+c_{1})}
\end{equation}
Integrating equation (\ref{eq27}), we obtain
\begin{equation}
\label{eq28}
\beta = k_{1}\left(n\ell t+c_{1}\right)^{-\frac{3}{n}}
\end{equation}
where $k_{1}$ is the constant of integration.\\

The expressions for the isotropic pressure ($p$), the proper energy
density ($\rho$), the string tension ($\lambda$) and the particle
density ($\rho_{p}$) for the above model are obtained as
\begin{eqnarray}\label{eq29}
  p &=& \frac{3\ell^{2}[n(m+1)- 3(m^{2}-m+1)]}{(m + 1)^{2}}
(n \ell t + c_{1})^{-2}+\frac{3(m+1)(n+1)}{4(9m-3+mn+n)}(n\ell t +
c_{1})^{-\frac{6(1 - m)} {n(m + 1)}} \nonumber\\
& & +\frac{3c_{3}\ell(1-2m)}{m+1}(n\ell t +c_{1})^{-\frac{3}{n}-1}-
\frac{(m+1)^{2}}{\ell^{2}(9m-3+mn+n)^{2}}(n\ell t +
c_{1})^{-\left[\frac{12(1 - m)} {n(m + 1)}-2\right]}\nonumber\\
& &- \frac{2c_{3}(m+1)}{\ell(9m-3+mn+n)}(n\ell t +
c_{1})^{-\left[\frac{3(3 - m)} {n(m + 1)}-1\right]}- c_{3}^{2}(n\ell
t + c_{1})^{-\frac{6}{n}}-\frac{3}{4}k^{2}(n\ell t + c_{1})^{-\frac{6}{n}},
\end{eqnarray}
\begin{eqnarray}\label{eq30}
\rho &=& \frac{9\ell^{2}m(2-m)}{(m + 1)^{2}} (n \ell t +
c_{1})^{-2}+\frac{15-33m-mn-n}{4(9m-3+mn+n)}(n\ell t +
c_{1})^{-\frac{6(1 - m)} {n(m + 1)}}  \nonumber\\
& &+\frac{3c_{3}\ell(1-2m)}{m+1}(n\ell t +c_{1})^{-\frac{3}{n}-1}-
\frac{(m+1)^{2}}{\ell^{2}(9m-3+mn+n)^{2}}(n\ell t +
c_{1})^{-\left[\frac{12(1 - m)} {n(m + 1)}-2\right]} \nonumber\\
& &- \frac{2c_{3}(m+1)}{\ell(9m-3+mn+n)}(n\ell t+
c_{1})^{-\left[\frac{3(3 - m)} {n(m + 1)}-1\right]}- c_{3}^{2}(n\ell
t + c_{1})^{-\frac{6}{n}} + \frac{3}{4}k^{2}(n\ell t + c_{1})^{-\frac{6}{n}},
\end{eqnarray}

\begin{equation}
\label{eq31} \lambda = \frac{3\ell^{2}(2m-1)(3 - n)}{m + 1}(n\ell t
+ c_{1})^{-2} + 2(n\ell t + c_{1})^{-\frac{6(1 - m)}{n(m + 1)}} \;,
\end{equation}
\begin{eqnarray}\label{eq32}
\rho_{p} &=& \frac{3\ell^{2}[3m(2-m))+(3-n)(1-2m)(m+1)]}{(m +
1)^{2}} (n \ell t +
c_{1})^{-2}+\frac{3(13-35m-3mn-3n)}{4(9m-3+mn+n)}(n\ell t +
c_{1})^{-\frac{6(1 - m)} {n(m + 1)}}  \nonumber\\
& &+\frac{3c_{3}\ell(1-2m)}{m+1}(n\ell t +c_{1})^{-\frac{3}{n}-1}-
\frac{(m+1)^{2}}{\ell^{2}(9m-3+mn+n)^{2}}(n\ell t +
c_{1})^{-\left[\frac{12(1 - m)} {n(m + 1)}-2\right]} \nonumber\\
& &- \frac{2c_{3}(m+1)}{\ell(9m-3+mn+n)}(n\ell t+
c_{1})^{-\left[\frac{3(3 - m)} {n(m + 1)}-1\right]}- c_{3}^{2}(n\ell
t + c_{1})^{-\frac{6}{n}} + \frac{3}{4}k^{2}(n\ell t + c_{1})^{-\frac{6}{n}}.
\end{eqnarray}
The critical energy density $(\rho_{c})$ and density parameter $(\Omega)$ are given by 
\begin{equation}
\label{eq33}
\rho_{c} = 3\ell^{2}(n\ell t+c_{1})^{-2}
\end{equation}
\begin{eqnarray}
\label{eq34}
\Omega &=& \frac{3m(2-m)}{(m + 1)^{2}}+\frac{15-33m-mn-n}{12(9m-3+mn+n)\ell^{2}}(n\ell t +
c_{1})^{-\frac{6(1 - m)} {n(m + 1)}+2}  \nonumber\\
& &+\frac{c_{3}\ell(1-2m)}{\ell(m+1)}(n\ell t +c_{1})^{-\frac{3}{n}+1}-
\frac{(m+1)^{2}}{3\ell^{4}(9m-3+mn+n)^{2}}(n\ell t +
c_{1})^{-\left[\frac{12(1 - m)} {n(m + 1)}-4\right]} \nonumber\\
& &- \frac{2c_{3}(m+1)}{3\ell^{3}(9m-3+mn+n)}(n\ell t+
c_{1})^{-\left[\frac{3(3 - m)} {n(m + 1)}-3\right]}- \frac{c_{3}^{2}}{3\ell^{2}}(n\ell
t + c_{1})^{-\frac{6}{n}+2} + \frac{k^{2}}{4\ell^{2}}(n\ell t + c_{1})^{-\frac{6}{n}+2},
\end{eqnarray}

The average Hubble's parameter, expansion scalar, shear scalar spatial volume ($V$) 
and anisotropy parameter $(\bar{A})$spatial 
volume ($V$) and anisotropy parameter $(\bar{A})$ of the
model are, respectively given by
\begin{equation}
\label{eq35} H = \frac{1}{3}(H_{x}+H_{y}+H_{z})=\ell(n\ell t +
c_{1})^{-1} ,
\end{equation}
\begin{equation}
\label{eq36} \theta = 3H=3\ell(n\ell t + c_{1})^{-1} ,
\end{equation}
\begin{eqnarray}\label{eq37}
\sigma^{2} &=&
\frac{1}{6}\left[(H_{x}-H_{y})^{2}+(H_{y}-H_{z})^{2}+(H_{z}-H_{x})^{2}\right]\nonumber\\
&=&\frac{3\ell^{2}(2m-1)^{2}}{(m+1)^{2}} (n \ell t +
c_{1})^{-2}-\frac{15-33m-mn-n}{4(9m-3+mn+n)}(n\ell t +
c_{1})^{-\frac{6(1 - m)} {n(m + 1)}}  \nonumber\\
& &-\frac{3c_{3}\ell(1-2m)}{m+1}(n\ell t +c_{1})^{-\frac{3}{n}-1}+
\frac{(m+1)^{2}}{\ell^{2}(9m-3+mn+n)^{2}}(n\ell t +
c_{1})^{-\left[\frac{12(1 - m)} {n(m + 1)}-2\right]} \nonumber\\
& &+ \frac{2c_{3}(m+1)}{\ell(9m-3+mn+n)}(n\ell t+
c_{1})^{-\left[\frac{3(3 - m)} {n(m + 1)}-1\right]}+ c_{3}^{2}(n\ell
t + c_{1})^{-\frac{6}{n}}.
\end{eqnarray}
\begin{equation}
\label{eq38} V = ABC=(n\ell t + c_{1})^{\frac{3}{n}},
\end{equation}
\begin{eqnarray}\label{eq39}
\bar{A} &=& \frac{2\sigma^{2}}{3H^{2}}=\frac{2(2m-1)^{2}}{(m+1)^{2}}
-\frac{15-33m-mn-n}{6\ell ^{2}(9m-3+mn+n)}(n\ell t +
c_{1})^{-\frac{6(1 - m)} {n(m + 1)}+2}  \nonumber\\
& &-\frac{2c_{3}(1-2m)}{\ell(m+1)}(n\ell t +c_{1})^{-\frac{3}{n}+1}+
\frac{2(m+1)^{2}}{3\ell^{4}(9m-3+mn+n)^{2}}(n\ell t +
c_{1})^{-\left[\frac{12(1 - m)} {n(m + 1)}-4\right]} \nonumber\\
& &+ \frac{4c_{3}(m+1)}{3\ell^{3}(9m-3+mn+n)}(n\ell t+
c_{1})^{-\left[\frac{3(3 - m)} {n(m + 1)}-3\right]}+
\frac{2c_{3}^{2}}{3\ell^{2}}(n\ell t + c_{1})^{-\frac{6}{n}+2}.
\end{eqnarray}

%%%%%%%%%%%%%%%%%%%%%%%%%%%%%%%%%%%%%%%%%%%%%%%%%%%%%%%%%%%%%%%%%%%%
\begin{figure}
\begin{center}
\includegraphics [height=8 cm]{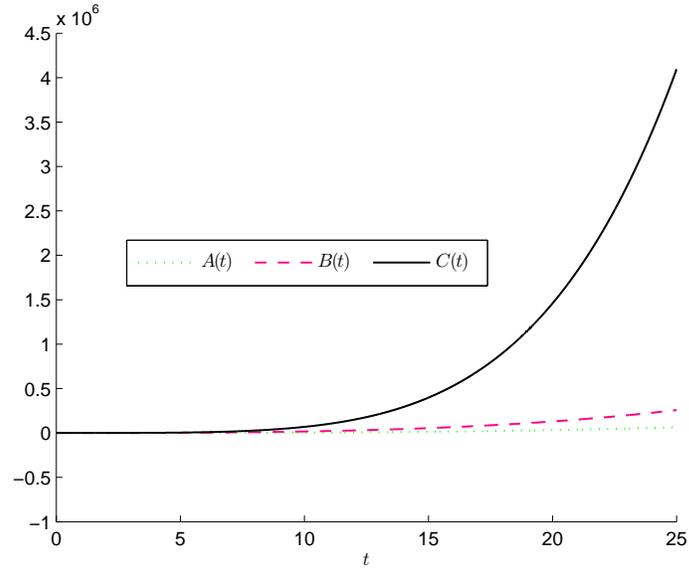}
\caption{Scale factors versus time (t).} \label{fg:anil33F1.eps}
\end{center}
\end{figure}
%%%%%%%%%%%%%%%%%%%%%%%%%%%%%%%%%%%%%%%%%%%%%%%%%%%%%%%%%%%%%%%%%%%
\begin{figure}
\begin{center}
\includegraphics [height=8 cm]{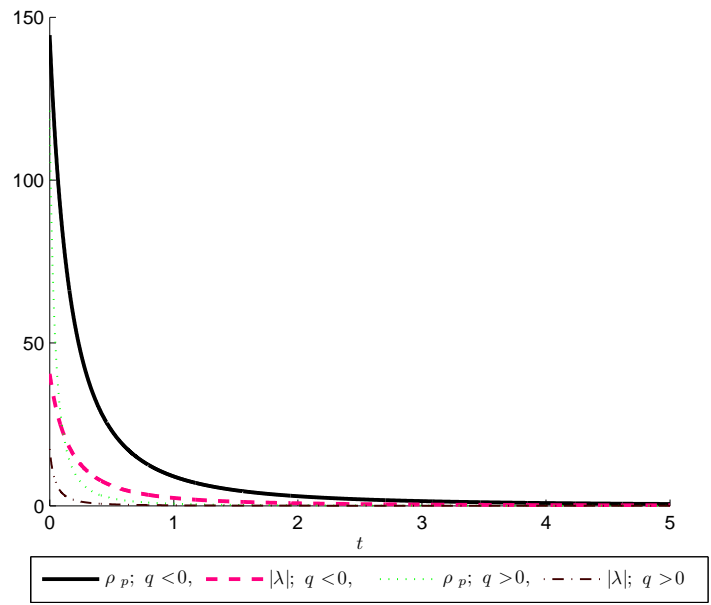}
\caption{Particle energy density $(\rho_{p})$ and string tension density $(\arrowvert \lambda \arrowvert)$ versus time (t).} 
\label{fg:anil33F2.eps}
\end{center}
\end{figure}
%%%%%%%%%%%%%%%%%%%%%%%%%%%%%%%%%%%%%%%%%%%%%%%%%%%%%%%%%%%%%%%%%%%%%%%%%
\begin{figure}
\begin{center}
\includegraphics [height=8 cm]{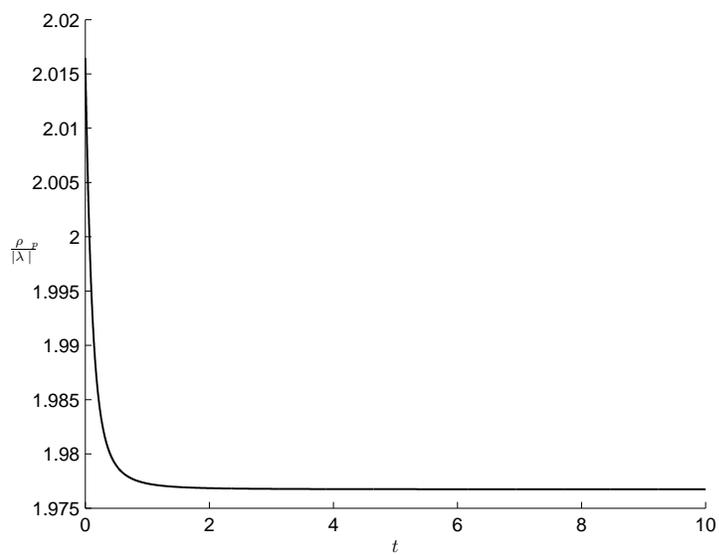}
\caption{Plot of $\frac{\rho_{p}}{\arrowvert \lambda \arrowvert}$ versus time (t).} \label{fg:anil33F3.eps}
\end{center}
\end{figure}
%%%%%%%%%%%%%%%%%%%%%%%%%%%%%%%%%%%%%%%%%%%%%%%%%%%%%%%%%%%%%%%%%%%%%%%%%%
\begin{figure}
\begin{center}
\includegraphics [height=8 cm]{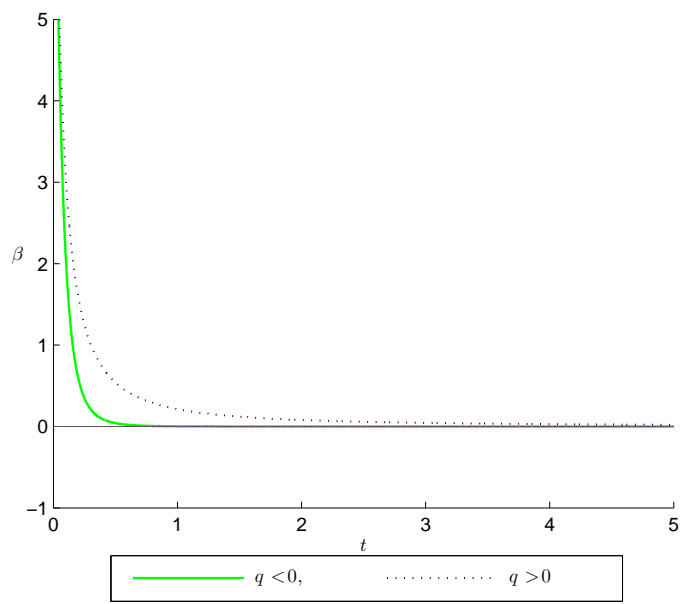}
\caption{Plot of displacement vector $(\beta)$ versus time (t).} \label{fg:anil33F4.eps}
\end{center}
\end{figure}
%%%%%%%%%%%%%%%%%%%%%%%%%%%%%%%%%%%%%%%%%%%%%%%%%%%%%%%%%%%%%%%%%%%%%%%%%%%%%%%%%
\begin{figure}
\begin{center}
\includegraphics [height=8 cm]{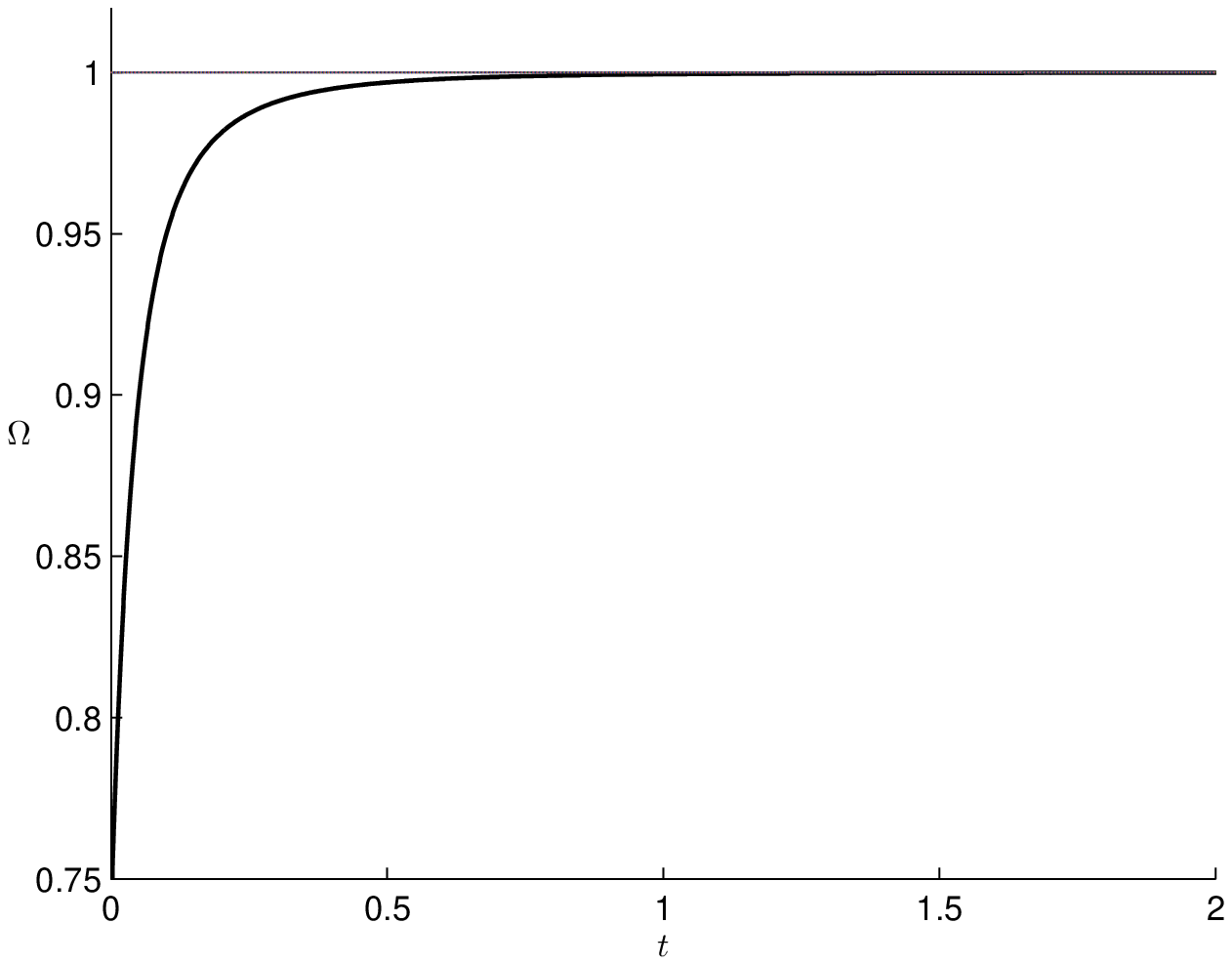}
\caption{Plot of density parameter $(\Omega)$ versus time (t).} \label{fg:anil323F5.eps}
\end{center}
\end{figure}
%%%%%%%%%%%%%%%%%%%%%%%%%%%%%%%%%%%%%%%%%%%%%%%%%%%%%%%%%%%%%%%%%%%%%%%%%%%%%%%%%%%%

We observe that all the parameters diverge at $t=-c_{1}/n\ell$.
Therefore, the model has a singularity at $t=-c_{1}/n\ell$, which
can be shifted to $t=0$ by choosing $c_{1}=0$. This singularity is
of Point Type as all the scale factors vanish at $t=-c_{1}/n\ell$.
The cosmological evolution of Bianchi-II space-time is expansionary
since all the scale factors monotonically increase with time (see,
$\textbf{Fig.1}$). So, the Universe starts expanding with a big bang
singularity in the derived model. The parameters $p$, $\rho$,
$\rho_{p}$ and $\lambda$ start off with extremely large values. In
particular, the large values of $\rho_{p}$ and $\lambda$ in the
beginning suggest that strings dominate the early Universe. For
sufficiently large times, $\rho_{p}$ and $\lambda$ become
negligible. Therefore, the strings disappear from the Universe for
larger times. \\

From $\textbf{Fig. 1}$, it is observed that during cosmic expansion, $C(t) > B(t) > A(t)$. Thus the 
derived model can be utilised to describe the anisotroy of early Universe. In particular for $m = 0.5$, at 
late time, the directional scale factors vary as 
\[A(t)\approx(n \ell t + c_{1})^{\frac{1}{n}},
\;\;B(t)\approx(n \ell t + c_{1})^{\frac{1}{n}},\;\;C(t)\approx(n
\ell t + c_{1})^{\frac{1}{n}}.\]
Therefore, isotropy is achieved in the derived model for $m=0.5$.\\
The value of DP ($q$) is found to be
\begin{equation}\label{40}
q=-1+\frac{\dot{H}}{H^{2}}=n-1,
\end{equation}
which is a constant. A positive sign of $q$, i.e., $n>1$ corresponds
to the standard decelerating model whereas the negative sign of $q$,
i.e., $0< n<1$ indicates acceleration. The expansion of the Universe
at a constant rate corresponds to $n=1$, i.e., $q=0$. Also, recent
observations of SN Ia \cite{ref43}-\cite{ref47} reveal that the present
Universe is accelerating and value of DP lies somewhere in the range
$-1<q< 0.$ It follows that in the derived model, one can choose the
values of DP consistent with the observations. 
For $n < 1$, the model is accelerating whereas for $n > 1$ it goes
to decelerating phase. In what follows, we compare the two modes of
evolution through graphical analysis of various parameters.\\

$\textbf{Fig. 2}$ depicts the variation of $\rho_{p}$ and $\arrowvert \lambda \arrowvert$ versus time 
in the decelerating and accelerating
modes of Universe. It is observed that in the early Universe $\rho_{p} > \arrowvert \lambda \arrowvert$, 
and for sufficiently large time,$\rho_{p}$ and $\arrowvert \lambda \arrowvert$ tend to zero. 
Therefore, the strings disappear from the Universe at late time (i. e. present epoch). 
According to Ref. \cite{ref17}, since there is no direct evidence of
strings in the present-day Universe, we are in general, interested
in constructing models of a Universe that evolves purely from the
era dominated by either geometric string or massive strings and ends
up in a particle dominated era with or without remnants of strings.
Therefore, the above model describes the evolution of the Universe
consistent with the present-day observations. $\frac{\rho_{p}}{\arrowvert \lambda \arrowvert}$ has been 
graphed in $\textbf{Fig. 3}$. It is evident that $\frac{\rho_{p}}{\arrowvert \lambda \arrowvert}$ is decreasing 
function of time and $\frac{\rho_{p}}{\arrowvert \lambda \arrowvert} > 1$ in the early Universe. This 
shows that massive strings dominate the early Universe. But the string tension density drops to zero for sufficiently 
large times. That is why the strings are not observed in current astronomical observations.\\

From equation (\ref{eq28}), it is clear that displacement vector $(\beta)$ 
is decreasing function of time and approaches to small positive value. This behaviour of $\beta$ is 
clearly shown in $\textbf{Fig. 4}$, in accelerating and decelerating phase of the Universe. From equation (\ref{eq34}), 
it is observed that for $n < 3$ and $m = 0.5$, the density parameter $(\Omega)$ approaches to 1 at late time. 
Thus the derived model predicts a flat Universe at the present epoch. $\textbf{Fig. 5}$ depicts the variation of density 
parameter versus time during the evolution of Universe.\\
   
\subsection{Lyra's Cosmology of String fluid with Exponential-law}
Solving the equations (\ref{eq17}), (\ref{eq12}) and (\ref{eq19}),
we obtain the metric functions as
\begin{equation}
\label{eq41} A(t) =
c_{2}^{\frac{3m}{m+1}}\exp{\left(\frac{3m\ell}{m+1}t\right)}\; ,
\end{equation}
\begin{equation}
\label{eq42} B(t) =
c_{4}c_{2}^{\frac{3m}{m+1}}\exp{\left[\frac{3m\ell}{m+1}t
+\frac{c_{2}^{\frac{6(m-1)}{m+1}}(m+1)^{2}}{18\ell^{2}(3m-1)(m-1)}e^{\frac{6\ell(m-1)}{m+1}t}-
\frac{c_{3}} {3\ell c_{2}^{3}}e^{-3\ell t}\right]},
\end{equation}
\begin{equation}
\label{eq43} C(t) =
c_{4}^{-1}c_{2}^{\frac{3(1-m)}{m+1}}\exp{\left[\frac{3(1-m)\ell}{m+1}t
-\frac{c_{2}^{\frac{6(m-1)}{m+1}}(m+1)^{2}}{18\ell^{2}(3m-1)(m-1)}e^{\frac{6\ell(m-1)}{m+1}t}+
\frac{c_{3}} {3\ell c_{2}^{3}}e^{-3\ell t}\right]}.
\end{equation}
Equation (\ref{eq16}) leads to
\begin{equation}
\label{eq44}
\frac{\dot{\beta}}{\beta}=-3\ell
\end{equation}
Integrating equation (\ref{eq35}), we obtain
\begin{equation}
\label{eq45}
\beta=k_{0} e^{-3\ell t}
\end{equation}
where $k_{0}$ is an integrating constant.\\

The expressions for the isotropic pressure, the proper energy
density, the string tension and the particle density for the derived
model are obtained as
\begin{eqnarray}
\label{eq46} p &=& -\frac{9\ell^{2}(m^{2}-m+1)}{(m + 1)^{2}}
+\frac{c_{2}^{\frac{6(m-1)}{m+1}}(m+1)}{4(3m-1)}e^{\frac{6\ell(m-1)}{m+1}t}
-\frac{2c_{3}c_{2}^{\frac{3(m-3)}{m+1}}(m+1)}{3\ell(3m-1)}e^{\frac{3\ell(m-3)}{m+1}t} 
\nonumber\\
& &+\frac{3c_{3}(1-2m)} {c_{2}^{3}(m+1)}e^{-3\ell t}
-\frac{c_{2}^{\frac{12(m-1)}{m+1}}(m+1)^{2}}{9\ell^{2}(3m-1)^{2}}e^{\frac{12\ell(m-1)}{m+1}t}
-\frac{c_{3}^{2}}{c_{2}^{6}}e^{-6\ell t} - \frac{3}{4}k_{0}^{2}e^{-3\ell t},
\end{eqnarray}
\begin{eqnarray}\label{eq47}
\rho &=& \frac{9\ell^{2}m(2-m)}{(m + 1)^{2}}
+\frac{c_{2}^{\frac{6(m-1)}{m+1}}(5-11m)}{4(3m-1)}e^{\frac{6\ell(m-1)}{m+1}t}
-\frac{2c_{3}c_{2}^{\frac{3(m-3)}{m+1}}(m+1)}{3\ell(3m-1)}e^{\frac{3\ell(m-3)}{m+1}t}
\nonumber\\
& &+\frac{3c_{3}(1-2m)} {c_{2}^{3}(m+1)}e^{-3\ell t}
-\frac{c_{2}^{\frac{12(m-1)}{m+1}}(m+1)^{2}}{9\ell^{2}(3m-1)^{2}}e^{\frac{12\ell(m-1)}{m+1}t}
-\frac{c_{3}^{2}}{c_{2}^{6}}e^{-6\ell t} + \frac{3}{4}k_{0}^{2}e^{-3\ell t}
\end{eqnarray}
\begin{equation}
\label{eq48} \lambda = \frac{9\ell^{2}(2m-1)}{m +
1}+2c_{2}^{\frac{6(m - 1)}{m + 1}} e^{\frac{6\ell(m-1)}{m+1}t},
\end{equation}
\begin{eqnarray}\label{eq49}
\rho_{p} &=& \frac{9\ell^{2}(1+m-3m^{2})}{(m + 1)^{2}}
+\frac{c_{2}^{\frac{6(m-1)}{m+1}}(13-25m)}{4(3m-1)}e^{\frac{6\ell(m-1)}{m+1}t}
-\frac{2c_{3}c_{2}^{\frac{3(m-3)}{m+1}}(m+1)}{3\ell(3m-1)}e^{\frac{3\ell(m-3)}{m+1}t}\nonumber\\
& &+\frac{3c_{3}(1-2m)} {c_{2}^{3}(m+1)}e^{-3\ell t}
-\frac{c_{2}^{\frac{12(m-1)}{m+1}}(m+1)^{2}}{9\ell^{2}(3m-1)^{2}}e^{\frac{12\ell(m-1)}{m+1}t}
-\frac{c_{3}^{2}}{c_{2}^{6}}e^{-6\ell t} + \frac{3}{4}k_{0}^{2}e^{-3\ell t}.
\end{eqnarray}

The critical energy density $(\rho_{c})$ and density parameter $(\Omega)$ are given by 
\begin{equation}
\label{eq50}
\rho_{c} = 3\ell^{2}
\end{equation}
\begin{eqnarray}\label{eq51}
\Omega &=& \frac{3m(2-m)}{(m + 1)^{2}}
+\frac{c_{2}^{\frac{6(m-1)}{m+1}}(5-11m)}{12\ell^{2}(3m-1)}e^{\frac{6\ell(m-1)}{m+1}t}
-\frac{2c_{3}c_{2}^{\frac{3(m-3)}{m+1}}(m+1)}{9\ell^{2}(3m-1)}e^{\frac{3\ell(m-3)}{m+1}t}
\nonumber\\
& &+\frac{3c_{3}(1-2m)} {3\ell^{2}c_{2}^{3}(m+1)}e^{-3\ell t}
-\frac{c_{2}^{\frac{12(m-1)}{m+1}}(m+1)^{2}}{27\ell^{4}(3m-1)^{2}}e^{\frac{12\ell(m-1)}{m+1}t}
-\frac{c_{3}^{2}}{3\ell^{2}c_{2}^{6}}e^{-6\ell t} + \frac{1}{4\ell^{2}}k_{0}^{2}e^{-3\ell t}
\end{eqnarray}

The average Hubble's parameter, expansion scalar, shear scalar spatial volume ($V$), 
 and anisotropy parameter $(\bar{A})$ of the
model are, respectively given by
\begin{equation}
\label{eq52} H = \ell.
\end{equation}
\begin{equation}
\label{eq53} \theta = 3 \ell,
\end{equation}
\begin{eqnarray}\label{eq54}
\sigma^{2} &=& \frac{3\ell^{2}(2m-1)^{2}}{(m + 1)^{2}}
-\frac{c_{2}^{\frac{6(m-1)}{m+1}}(5-11m)}{4(3m-1)}e^{\frac{6\ell(m-1)}{m+1}t}
+\frac{2c_{3}c_{2}^{\frac{3(m-3)}{m+1}}(m+1)}{3\ell(3m-1)}e^{\frac{3\ell(m-3)}{m+1}t}
\nonumber\\
& &-\frac{3c_{3}(1-2m)} {c_{2}^{3}(m+1)}e^{-3\ell t}
+\frac{c_{2}^{\frac{12(m-1)}{m+1}}(m+1)^{2}}{9\ell^{2}(3m-1)^{2}}e^{\frac{12\ell(m-1)}{m+1}t}
+\frac{c_{3}^{2}}{c_{2}^{6}}e^{-6\ell t},
\end{eqnarray}
\begin{equation}
\label{eq55} V = c_{2}^{3}e^{3\ell t},
\end{equation}
\begin{eqnarray}\label{eq56}
\bar{A} &=& \frac{2(2m-1)^{2}}{(m + 1)^{2}}
-\frac{c_{2}^{\frac{6(m-1)}{m+1}}(5-11m)}{6\ell^{2}(3m-1)}e^{\frac{6\ell(m-1)}{m+1}t}
+\frac{4c_{3}c_{2}^{\frac{3(m-3)}{m+1}}(m+1)}{9\ell^{3}(3m-1)}e^{\frac{3\ell(m-3)}{m+1}t}
\nonumber\\
& &-\frac{2c_{3}(1-2m)} {\ell^{2}c_{2}^{3}(m+1)}e^{-3\ell t}
+\frac{2c_{2}^{\frac{12(m-1)}{m+1}}(m+1)^{2}}{27\ell^{4}(3m-1)^{2}}e^{\frac{12\ell(m-1)}{m+1}t}
+\frac{2c_{3}^{2}}{3\ell^{2}c_{2}^{6}}e^{-6\ell t},
\end{eqnarray}

The value of DP ($q$) is found to be
\begin{equation}
\label{eq57} q = - 1.
\end{equation}

Recent observations of SN Ia \cite{ref43}$-$\cite{ref47} suggest that the
Universe is accelerating in its present state of evolution. It is
believed that the way Universe is accelerating presently; it will
expand at the fastest possible rate in future and forever. For
$n=0$, we get \textbf{ $q=-1$ }; incidentally this value of DP leads
to $dH/dt=0$, which implies the greatest value of Hubble's parameter
and the fastest rate of expansion of the Universe. Therefore, the
derived model can be utilized to describe the dynamics of the late
time evolution of the actual Universe. From equation (\ref{eq51}), it is clear that the density parameter 
($\Omega$) approaches to 1 at late time. Thus the derived model predicts a flat Universe.

\section{Concluding Remarks}
In this paper, we have studied Bianchi-II string cosmological models in normal gauge for Lyra's manifold with 
constant deceleration parameter, considering two cases, ($3.1$) and ($3.2$). for $n \neq 0$ and $n = 0$ respectively. 
It is observed that in both cases, the mean anisotropy parameter ($\bar{A}$) drops to zero for $m=0.5$ 
for sufficiently large times. Thus for 
$m = 0.5$, isotropy is achieved in derived models at late time. The main features of the work are as follows:\\ 
  
\begin{itemize}
\item The models are based on exact solutions of modified Einstein's field equations in normal gauge for 
Lyra's manifold for the anisotropic Bianchi-II
space-time filled with massive strings.
\item The singular model ($n\neq 0$) seems to describe the dynamics of
    Universe from big bang to the present epoch while the
    non-singular model ($n=0$) seems reasonable to project dynamics of
    future Universe.
\item The Universe acquires flatness at late time (see $\textbf{Fig. 5}$) which is consistent with the prediction 
of current observations.
\item The displacement vector $(\beta)$ is a decreasing function of time and it approaches to small 
positive value at late time. This matches with the nature of cosmological constant $(\Lambda)$.  
\item  The string tension density vanishes for sufficiently large times (see $\textbf{Fig. 2}$). Thus the strings 
disappear from Universe at late time. That is why, the strings are not observable in the present Universe. 
\item The age of Universe in singular model is given by
$$T_{0}=\frac{1}{n}H_{0}^{-1}-\frac{c_{1}}{n\ell}$$
which differ from present estimation i. e. $T_{0} = H_{0}^{-1} \approx 14 Gyr$. But 
if we take, $n = 1$ and $c_{1} = 0$, the model is in good agreement with present age of Universe. 

\end{itemize}

\end{document}